\pdfoutput=1
\documentclass[sigconf]{acmart}
\acmConference[EASE 2025]{The 29th International Conference on Evaluation and Assessment in Software Engineering}{17–20 June, 2025}{Istanbul, Türkiye}

\usepackage{graphicx}
\usepackage{pgfplots}
\usetikzlibrary{patterns}
\pgfplotsset{compat=1.18}
\usepackage{xcolor}
\usepackage{hyperref} 
\hypersetup{
    colorlinks=true,
    linkcolor=blue,
    filecolor=blue,      
    urlcolor=blue,
    citecolor=blue,
    pdftitle={A Survey for What Developers Require in AI-powered Tools that Aid in Component Selection in CBSD}
}
\usepackage{xspace} 
\usepackage{multirow}
\usepackage{comment}

\newcommand{\modified}[1]{{\color{black}{#1}}}

\newcommand{\tblname}[1]{{\color{black}{#1}}}

\newcommand{\inlinequote}[1]{\textit{``#1''}} 
\newcommand{\inlinename}[1]{`#1'} 
\newcommand{\inlinecode}[1]{{\fontfamily{lmdh}\selectfont{\small\textsl{#1}}\normalfont}}

\newcommand{\num}[1]{\textcolor{black}{#1}}

\usepackage[T1]{fontenc}
\usepackage[utf8]{inputenc}
\usepackage{enumitem}
\usepackage{tabularx}
\newcolumntype{Y}{>{\raggedright\arraybackslash}X} 
\usepackage{longtable}
\usepackage{makecell}
\usepackage{ltablex}
\usepackage{array}
\usepackage{booktabs}
\usepackage{ragged2e}
\newcolumntype{L}[1]{>{\raggedright\arraybackslash}p{#1}}
\newcolumntype{C}[1]{>{\centering\arraybackslash}m{#1}}
\usepackage[normalem]{ulem}
\usepackage{amsmath}
\makeatletter
\newcommand\footnoteref[1]{\protected@xdef\@thefnmark{\ref{#1}}\@footnotemark}
\makeatother

\begin{document}

\title{A Survey for What Developers Require in AI-powered Tools that Aid in Component Selection in CBSD}

\author{Mahdi Jaberzadeh Ansari}
\email{mahdi.ansari1@ucalgary.ca}
\affiliation{%
  \institution{Schulich School of Engineering, University of Calgary}
  \streetaddress{ICT 402, 2500 University Drive NW}
  \city{Calgary}
  \state{AB}
  \country{Canada}
  \postcode{T2N 1N4}
} 

\author{Ann Barcomb}
\email{ann@barcomb.org}
\affiliation{%
  \institution{Schulich School of Engineering, University of Calgary}
  \streetaddress{ICT 402, 2500 University Drive NW}
  \city{Calgary}
  \state{AB}
  \country{Canada}
  \postcode{T2N 1N4}
} 

\begin{abstract}
Although it has been more than \num{four} decades that the first components-based software \modified{development (CBSD)} studies were conducted, there is still no standard method or tool for component selection which is widely accepted by the industry. The gulf between industry and academia contributes to the lack of an accepted tool. We conducted a mixed methods survey of nearly \num{100} people engaged in component-based software engineering practice or research to better understand the problems facing industry, how these needs could be addressed, and current best practices employed in component selection. We also sought to identify and prioritize quality criteria for component selection from an industry perspective. In response to the call for \modified{CBSD} component selection tools to incorporate recent technical advances, we also explored the perceptions of professionals about AI-driven tools, present and envisioned.
\end{abstract}



\maketitle

\section{Introduction}\label{sec:introduction}

The core concept of component-based software development (CBSD) is the reuse of ready-to-use software blocks known as components~\cite{khan2021critical}. Reusing existing components significantly reduces development costs, time to market, and various risks, such as security vulnerabilities~\cite{khan2021critical}. Software development based on existing components is more popular than developing software from scratch~\cite{fatima2017risk, khan2021critical}. Many software companies adopt CBSD because it not only reduces costs and shortens development time, but also indirectly provides access to highly skilled professionals.
Furthermore, CBSD is modular, which allows developers to replace components without upgrading the entire system~\cite{khan2021critical}.

The benefits of \modified{CBSD} are recognized by industry; however, the application of formal \modified{CBSD} is still limited to very specific domains, such as avionics, railways, and automotive software, which require high standards of safety and reliability~\cite{li2023challenges}. There are still many gaps in our understanding of CBSE, which act as barriers to a more effective and broader application of \modified{CBSD} in real-world projects. \citeauthor{vale2016twenty} (\citeyear{vale2016twenty}) conducted a thorough systematic mapping study of the literature published between 1984 and 2012. They reported \num{nine} research gaps. These include the lack of validation of CBSD methods in industrial settings, the lack of a standard set of quality criteria for component selection, and the lack of software tools to support developers to implement \modified{CBSD} approaches effectively~\cite{vale2016twenty}. 
\modified{More recent studies have identified the same limitations~\cite{petersen2017choosing, badampudi2018decision}. For example, a }
recent literature review by \citeauthor{mahapatro2024reviewing} (\citeyear{mahapatro2024reviewing}) reported that the lack of automated software tools to support component selection remains an issue~\cite{mahapatro2024reviewing}. The integration of machine learning (ML) and artificial intelligence (AI) in \modified{CBSD} techniques is also lacking~\cite{mahapatro2024reviewing}. 
Many component selection processes applied in industry have been observed to be suboptimal, demonstrating the need for better support to improve decision-making outcomes~\cite{petersen2017choosing}.

\citeauthor{storey2020software} (\citeyear{storey2020software}) highlighted that while many software engineering (SE) studies aim to benefit human stakeholders, they often focus predominantly on technical contributions without engaging developers in the research process. When research is not conducted in collaboration with industry, the resulting research may not fit the context and may not meet industry needs~\cite{wohlin2013empirical}.
Only \num{52} out of \num{1231} \modified{CBSD} studies over \num{28} years were based on industrial experience (\num{4\%}) 
\cite{vale2016twenty}. 
The majority of component selection tools are obsolete or unavailable, 
while the majority of proposed algorithms have 
not been adopted by industry or received further attention from researchers
\cite{lenarduzzi2020open,konys2013approach}, suggesting engaging with practitioners presents a significant opportunity to aid researchers in developing relevant and useful tools to support component selection.

We were motivated to conduct a mixed methods survey to better understand developers' needs in an \modified{AI-driven} component selection tool because current methods and tools are inconvenient and do not meet developers' needs, 
and engagement with practitioners is critical to developing relevant and useful solutions.
\modified{In response to the limited number of CBSD studies in industry settings,} this survey aims to explore and understand key aspects of software component selection in industry and the potential role of AI in improving these processes. \modified{In contrast to recent studies, we focused primarily on the potential of AI in CBSD, which is an identified gap~\cite{mahapatro2024reviewing}.} Therefore, we define the following research questions:

\begin{enumerate}[label=\textbf{RQ\arabic*)}, widest=RQ7, align=left]
    \item\label{RQ1} What are the most frequently used methods for component selection in practice?
    \item\label{RQ2} What tools do practitioners rely on for component selection?
    \item\label{RQ3} What quality criteria are prioritized in the industry for the selection of components?
    \item\label{RQ4} What best practices are followed in the industry for the successful selection of components?    
    \item\label{RQ5} How do professionals perceive AI-driven tools for component selection, particularly in terms of barriers, enablers, adoption, integration, and intended features?
\end{enumerate}

The remainder of this paper is organized as follows.
Section~\ref{sec:related work} discusses previous studies in CBSE, including surveys, interviews, and experiments that capture industry perspectives.
Section~\ref{sec:methodology} explains the research methods used for data collection and analysis. Section~\ref{sec:results} provides descriptive statistics and presents the result of qualitative data analysis.
Section~\ref{sec:discussion} directly addresses the research questions, discusses the implications of the results, highlights the immediate needs of the industry. Section~\ref{sec:limitations} outlines the limitations of this study.
Finally, Section~\ref{sec:conclusion} summarizes the key findings of our research.

\section{Related Work}\label{sec:related work}

This section reviews recent studies about component selection for \modified{CBSD} in industry. 
We cover how companies are currently selecting components, the gap between research and practice, and the software attributes that inform the decision process. Due to the aforementioned disconnect between industry and academia, there has been limited research addressing these topics.
\modified{For this reason, we have also considered studies on how practitioners select third-party libraries, even if the studies do not explicitly focus on CBSD.}

Developers tend to use informal methods in their component selection, such as expert judgment, manual data collection, and search engines rather than systemic and tool-based approaches~\cite{chen2007survey, petersen2017choosing, ayala2011selection}. Reliance on informal methods, difficulties in adapting components to requirements, and lack of standardized tools often lead to suboptimal results~\cite{li2005empirical, petersen2017choosing, larios2020selecting}, costing both time and money~\cite{farshidi2018decision, badampudi2018decision, siena2014modelling}. The evaluation of components can be improved through unit testing, communication and training within the team, and early planning~\cite{ilyas2017empirical}.

Academic proposals are often misaligned with industry needs. Although companies are interested in evaluation tools, there has been minimal adoption of formal selection methods~\cite{ayala2011selection, petersen2017choosing}. Most traditional component selection methods---such as multi-criteria decision-making models---lack scalability, making them costly and impractical for large or evolving criteria~\cite{farshidi2018decision}. Formalized methods are challenging not only because of diverse client needs and varying project domains, but also because of opportunistic reuse practices which means that a decision made in one context might be deployed in a different context~\cite{ayala2011selection}. Practices of evaluating functional and non-functional attributes, adopting an iterative process linked with requirements engineering, assessing component combinations for architectural fit and total cost of the system, and prioritizing critical components with appropriate resource allocation and confidence levels could help companies design formal selection procedures tailored to their needs~\cite{land2008cots}. Effective component selection through formal methods could also be supported through validated, automated, up-to-date tools~\cite{farshidi2018decision, badampudi2018decision, siena2014modelling}.

The gap between industry requirements and academic research has led to multiple calls to involve practitioners as stakeholders in research~\cite{ayala2011selection, petersen2017choosing, badampudi2018decision}. This is a recurring concern not only for component selection, but also in several other areas of SE, such as software testing \cite{santos2019mind}. We are motivated to support companies in component selection with accessible and adoptable recommendations, using the approach of surveying practitioners, following examples such as \citeauthor{harutyunyan2019industry}'s (\citeyear{harutyunyan2019industry}) analysis of what companies need in a tool to govern the use of open source software in commercial products~\cite{harutyunyan2019industry} and \citeauthor{holstein2019improving}'s (\citeyear{holstein2019improving}) study on practitioner's needs regarding fairness testing techniques and tools~\cite{holstein2019improving}.

Practitioners prioritize different quality attributes according to their organizational needs. \modified{For example, developers rely on factors such as popularity, documentation, maintenance, and community support when selecting libraries \cite{larios2020selecting, nadi2023selecting}.} However, cost, support, longevity, functional suitability, and reliability are important factors in component selection \cite{chatzipetrou2018component, borg2019selecting, badampudi2018decision}. Smaller companies emphasize ease of development and maintainability, including complexity, API adequacy, and accessibility of documentation, while larger companies focus mainly on cost~\cite{chatzipetrou2018component, chatzipetrou2020component}.

\section{Methodology}\label{sec:methodology}

Surveys are a recognized research method for collecting data from the developer population~\cite{kitchenham2008personal, 
shull2008guide}. Although other research methods (e.g., case studies, experiments, or Delphi studies) can also draw deep insights into developer needs, surveys are particularly well suited to explore a broad set of experiences and generalized perceptions from a relatively larger number of participants~\cite{shull2008guide}. Therefore, we conducted an online survey to gather opinions from people  in the software industry. We used mixed methods because we wanted to not only capture the frequency of specific requirements or selection challenges, but also to gather written feedback on how developers approach and evaluate different components. This research was conducted with the ethical approval of the Ethical Review Board of our institution.

    \subsection{Survey Design}\label{sub:methodology:design}

We designed the survey to adapt to the responses of each participant, resulting in \num{20} to \num{40} possible questions. Only two questions were mandatory: one for consent at the beginning and one in the middle asking whether the respondent would consider using AI-driven tools in their component selection processes. If they answered \inlinequote{no,} the survey ended after a few additional questions. If they answered \inlinequote{yes,} additional questions explored their thoughts on AI-driven component selection. Before releasing the survey, we conducted a pilot test with two participants, and used their feedback to improve the survey. We organized the survey into \num{ten} sections, each focusing on a different aspect of component selection and AI adoption\modified{, with the number of questions per section indicated in curly braces}:

\begin{enumerate}
    \item \textbf{Demographics \modified{\{6\}}:}
    The career backgrounds of the participants, years of experience and geographic location were captured to contextualize their perspectives.
     
    \item \textbf{Warm-up \modified{\{1\}}:}
    We presented an academic definition of a software component and invited respondents to share their own understanding, ensuring a shared conceptual foundation.
     
    \item \textbf{Current Practices and Challenges \modified{\{8\}}:}
    The participants were asked about the existing component selection methods in their organization, including the quality attributes they measure, how they prioritize and evaluate these attributes, and the challenges they have faced.
     
    \item \textbf{Perceptions of AI-driven Tools \modified{\{3\}}:}
    We investigated whether participants currently use any tools for component selection and, if so, whether those tools incorporate AI features.
     
    \item \textbf{Desired Features and Functionality \modified{\{3\}}:}
    We explored the capabilities that practitioners believe are essential in a tool that supports component selection, inviting open-ended suggestions for potential enhancements.
     
    \item \textbf{Common Mistakes in Component Selection \modified{\{7\}}:}
    We encouraged participants to share their experiences of pitfalls, errors, and unintended consequences arising from their component selection processes.
     
    \item \textbf{Role of AI in Mitigating Mistakes \modified{\{3\}}:}
    We examined perceptions and experiences about the use of AI to address, reduce, or prevent mistakes reported in the previous section.
     
    \item \textbf{Challenges with AI tools \modified{\{2\}}:}
    This section focused on limitations of AI in component selection from the industry point of view and effective strategies for integrating AI into component selection.
     
    \item \textbf{Development Considerations \modified{\{3\}}:}
    We inquired about any experience the participants had in developing AI-driven selection tools, including details on model training approaches, data sources, and technical challenges.
     
    \item \textbf{Adoption and Integration \modified{\{3\}}:}
    Finally, we investigated the barriers and concerns around the integration of AI tools into existing workflows, exploring how such tools can be smoothly integrated into current development practices.
\end{enumerate}

    \subsection{Data Collection}\label{sub:methodology:collection}
        \subsubsection{Sampling Approach}\label{sub-sub-sec-sampling-approach}

\modified{As part of our inclusion criteria the} population of interest was people with experience in CBSD, including engineers, architects, project managers, and academic researchers in software engineering. 
\modified{We excluded people outside of this population.}

Due to the investigatory focus of this study and the accessibility of potential participants, we used a convenience sampling strategy to recruit individuals who had experience in component selection by soliciting responses from our personal networks. This method is \modified{consistent with \citeauthor{baltes2022sampling}'s (\citeyear{baltes2022sampling}) research on sampling strategies in SE} and is acknowledged to introduce biases and limit the generalization of the findings~\cite{baltes2022sampling}. We attempted to mitigate these concerns by also contacting people unknown to us through targeted emails to \modified{CBSD} scholars, LinkedIn InMail messages to software developers, posting invitations in software-related groups on LinkedIn, and encouraging snowball recruiting.

\subsubsection{Implementation of Data Collection}
The data was collected via an online questionnaire, hosted on the secure Qualtrics survey platform. The survey was available for \num{six} weeks from November 15, 2024, to December 31, 2024, to allow for broad participation and to accommodate varying participant schedules. After removing the records of the participants who left the survey empty or were flagged as duplicates or bots by Qualtrics algorithms, we reached \num{99} responses. A combination of multiple choice, ordering, and open-ended text fields was used to capture quantitative data on participant background and tool requirements, relative importance of various selection factors, and qualitative feedback, respectively. 

\subsubsection{Data Availability}
We share the following artifacts as part of our replication package \footnote{\url{https://figshare.com/s/7f79f26d2c1061387f53}}: ethics approval, recruitment emails, consent forms, survey questions, anonymous survey responses, the final codebook (in mx24 format), and the coded qualitative responses (in mtr24 format).

    \subsection{Data Analysis}\label{sub:methodology:qda}
        \subsubsection{Quantitative Data Analysis}
Quantitative responses (multiple choice, checkbox, and ordering items) were extracted from the survey platform into a database. Descriptive statistics such as frequency distributions were calculated to summarize findings on tool requirements, and basic comparisons \modified{(e.g., Chi-square, Mann–Whitney, Spearman, and Kruskal–Wallis tests)} were performed where relevant to assess differences between respondents. Lastly, we performed basic statistical analyzes to explore possible links between demographic data and the responses of the participants. 

\subsubsection{Qualitative Data Analysis (QDA)}
To analyze responses to open-text questions, we used a thematic qualitative data analysis approach~\cite{cruzes2011recommended, braun2006using}. First, anonymized participant comments were exported to separate text files corresponding to each respondent. We then imported text files into MaxQDA{\footnote{\url{https://www.maxqda.com/}}}. 

The first author coded the free-form text answers of \num{two} survey respondents and shared the coding schema with the second author. \modified{ For example, we applied \inlinecode{Question Refinement} (references to rewriting or clarifying questions) and \inlinecode{Adaptive Prompting} (adjusting prompts in real time based on feedback) to \inlinequote{Rewriting the questions/prompts} from R39 in response to Q31.} 

We checked the inter-rater agreement on the two documents and repeated the process of comparing two documents at a time until agreement exceeded \num{80\%}, which took two iterations. MaxQDA calculates agreement using Cohen's Kappa \cite{cohen:1960:coefficient}. Then the first author applied the finalized coding schema to the remaining documents. During this process we continued to check the effect of the evolution of the codebook on inter-rater reliability by having the second author code additional documents and discussing disagreements. The final codebook consisted of \num{324} codes and \num{2298} codings across \num{78} survey responses. \modified{Due to space limitations, we have not reported all themes.  Please refer to the replication package to see the complete coding schema.}

\section{Results}\label{sec:results}

We sent nearly \num{500} targeted emails and launched a LinkedIn campaign 
to invite a wider audience to participate in our study. A total of \num{118} individuals participated, although \num{19} responses were excluded because they were duplicates or empty submissions.  
\modified{Participants were asked to respond based on their own past experience.} The survey was adaptive, and participants were not shown questions which were not relevant to them.  All questions were optional as well, meaning that participants might have skipped some questions. For this reason, we report the number of responses \modified{to the number of times seen} per question in the format (Q\#, N=?\modified{/?})~\cite{kitchenham2008personal}. \num{Eighty-five} out of \num{99} participants \modified{reached the last page of the survey}, which Qualitrics calculated as a \num{92\%} response rate. \modified{Furthermore, participation in written questions (e.g., Q15, Q25, Q27, Q31–Q34) was relatively lower compared to other types of questions.}

In this paper, we use the following conventions to discuss the survey.
We use italics and double quotation marks to indicate a participant quotation, a literal or paraphrased question in our survey, and  a literal or paraphrased answer used as scaling options in matrix questions in our survey (e.g., \inlinequote{Very important}). Participant quotations are also identified by respondent number (R\#). Single quotation marks indicate predefined answer choices in multiple- or single-choice questions (e.g., \inlinename{ease of use}). A script font is used to indicate codes applied during QDA (e.g., \inlinecode{benchmarking}).

    \subsection{Demographics}

The experience of participants covers a wide range (Q1, N=\num{91/99}),
with \num{20} having less than \num{five} years of experience, \num{39} between \num{five} and \num{ten} years, and \num{32} over \num{ten} years.

The participants worked in a variety of sectors (Q2, N=\num{99/99}) and many indicated experience in more than one sector, which is why the sum of sectors exceeds the number of participants. IT and technology  had the highest number of participants (\num{71}), followed by finance (\num{25}) and education (\num{22}). Healthcare and pharmaceuticals  (\num{18}),  energy (\num{14}), and manufacturing and automotive (\num{12}) were also well-represented. A smaller number of participants worked in governments, public services, and non-profits (\num{9}), transportation and logistics (\num{6}), or   \inlinequote{Other} (\num{14}).

For education (Q3, N=\num{99/99}), \num{48} had a bachelor's degree in computer science or a related field,  \num{46} had a master's degree, and \num{14} had a Ph.D. A further \num{37} were self-taught, while \num{eight} attended coding boot camps. Trade or technical school (\num{2}) or other forms of education (\num{3}) were less common. \modified{Some} chose multiple options.

All participants shared their current roles (Q4, N=\num{99/99}), with \num{27} working as software developers and \num{23} as researchers in academic environments. Project managers and team leaders each accounted for \num{nine} participants, while \num{eight} were DevOps engineers and \num{seven} were software architects. IT managers formed a smaller group with \num{five} participants. Furthermore, \num{eleven} participants selected \inlinequote{Other,} and identified roles such as quality assurance engineer, product owner, CEO, consultant, network engineer, cloud engineer, and data engineer. \modified{Each person could report multiple roles.}

Most participants provided their organization size (Q6, N=\num{95/99}). The largest group, \num{23} participants, were working in organizations with more than 1001 employees. Medium organizations (1 to 10, 11 to 50 and 51 to 250 employees) are equally represented, each with \num{20} participants. Only \num{12} participants worked in organizations with 251 to 1000 employees.

The participants lived in various countries (Q5, N=\num{92/99}), with the highest number from Canada (\num{28}), followed by Germany (\num{18}) and Iran (\num{12}). Other countries with multiple participants were Armenia (\num{9}), USA (\num{6}), the Netherlands (\num{5}), UK (\num{3}), Italy (\num{3}), Pakistan (\num{3}), and India (\num{2}). The following countries each had one participant: Australia, China, Denmark, Finland, South Africa, and Switzerland. \modified{Due to our approved ethics, only limited demographic information was collected to maintain participant anonymity.}

    \subsection{Warm-up}

All participants revealed how familiar they were with AI technologies in general (Q7, N=\num{99/99}). The majority, \num{35}, identified themselves as \inlinequote{moderately familiar,} followed by \num{31} who were \inlinequote{very familiar.} \num{Seventeen} participants reported being \inlinequote{extremely familiar,} while \num{14} were \inlinequote{slightly familiar.} \num{Two} participants stated that they had \inlinequote{no familiarity.} 

We presented the definition of the software component provided by \citeauthor{kumar2015neuro} (\citeyear{kumar2015neuro}): {\inlinequote{A software component is a non-trivial, independently developed software package which provides services through specified interfaces and can be deployed independently. A component interacts with other component by using interfaces which can be connected with other components to compose a larger system.}~\cite[p. 1]{kumar2015neuro}} and asked them to express their agreement with the provided definition or share their own definition (Q8, N=\num{98/99}). The majority of participants (\num{94}) believed that this definition matched with what they knew as a software component, while \num{four} provided their own definition. Participants R25, a researcher with \num{ten} years of experience, said \inlinequote{The independent angle is not a hard definition for us,} suggesting that even packages injected into applications, such as NPM packages in Node.js or modules in React.js, can be considered components. This aligns with our observation that there is no single clear definition in the \modified{CBSD} literature.

    \subsection{Current Practices and Challenges}

All participants shared their preferred resources for collecting information about software components (Q9, N=\num{99/99}). \num{Eighty} used \inlinename{search engines and generative AI tools} such as Google or ChatGPT. \inlinename{Source and package repositories,} including GitHub and NPM, were used by \num{70} participants. Getting suggestions from \inlinename{colleagues, mentors, and developer meetups} were selected by \num{60}, while \num{51} participants each relied on \inlinename{online videos and tutorials}, \inlinename{Q\&A platforms} such as StackOverflow, and \inlinename{technical blogs.} \inlinename{Official documentation or vendor websites} were used by \num{46} participants, followed by \inlinename{internal organizational resources} (\num{37}), and \inlinename{conferences or industry events} (\num{33}). \inlinename{Social media platforms,} such as LinkedIn or Reddit, were mentioned by \num{29}. 

Participants were asked to identify the \num{five} most important criteria for evaluating software components (Q10, N=\num{98/99}). We then asked participants to rank the selected criteria from the most important (\num{first} position) to the least important (\num{fifth} position) (Q11, N=\num{78/99}). \tblname{Table \ref{tab:criteria_priorities}} presents the number of people who described the criterion as \inlinequote{important} (Imp.) and how many people ranked each criterion in one of the top five positions. Only the most important criteria are included; importance was calculated by giving {\num{five}} points for each first ranking, {\num{four}} points for each second ranking, and so on. Only criteria with a score of at least \num{24} are shown.

We provided an open-ended question for participants to describe their current process (Q12, N=\num{53/95}). Most rely on \inlinecode{ad hoc and manual methods} (\num{44}), followed by \inlinecode{test-driven approaches} (\num{26}), \inlinecode{sequential processes} (\num{14}), \inlinecode{market analysis} (\num{9}), \inlinecode{peer code reviews} (\num{8}), \inlinecode{requirement analysis} (\num{8}), \inlinecode{development of proof-of-concept} (\num{8}), \inlinecode{benchmarking} (\num{6}), and \inlinecode{testing demos or trials} (\num{6}). We also tried to identify quality criteria which were not included as an option in Q10. Participants mentioned  \inlinecode{return on investment}, \inlinecode{long-term viability}, \inlinecode{source code availability}, and \inlinecode{independent deployment} as additional quality criteria.

\renewcommand{\arraystretch}{1} 
\begin{table}[htbp]
    \centering
    \caption{Criteria for Evaluating Components: Number of People who Found Criterion Important (Imp.) and Number who Ranked it by Position. \modified{Ordered by \# of votes (Imp.).}}
    \label{tab:criteria_priorities}
    \begin{tabular}{L{0.37\linewidth}cccccc}
    
        \hline
        \textbf{Criterion} & \textbf{Imp.} &\textbf{1st} & \textbf{2nd} & \textbf{3rd} & \textbf{4th} & \textbf{5th} \\ \hline

        Reliability & \num{33} & \num{6} & \num{8} & \num{3} & \num{7} & \num{2} \\ 

        Active development or community activity & \num{30} & \num{9} & \num{4} & \num{5} & \num{5} & \num{1} \\ 

        Documentation quality & \num{29} & \num{3} & \num{7} & \num{3} & \num{4} & \num{7} \\ 
        
        Security & \num{27} & \num{6} & \num{8} & \num{6} & \num{2} & \num{3} \\ 
        
        Cost & \num{27} & \num{5} & \num{6} & \num{5} & \num{5} & \num{3} \\ 

        Ease of use & \num{27} & \num{4} & \num{4} & \num{4} & \num{4} & \num{5} \\ 

        Efficiency & \num{24} & \num{2} & \num{2} & \num{6} & \num{5} & \num{3} \\ 
        
        Architecture &  \num{22} & \num{9} & \num{3} & \num{1} & \num{3} & \num{2} \\ 

        Adaptability &  \num{19} & \num{1} & \num{0} & \num{3} & \num{4} & \num{3} \\ 
        
        Scalability &  \num{18} & \num{0} & \num{6} & \num{5} & \num{4} & \num{2} \\ 
        
        Functionality &  \num{17} & \num{8} & \num{4} & \num{2} & \num{1} & \num{0} \\ 

        License &  \num{15} & \num{7} & \num{3} & \num{2} & \num{2} & \num{1} \\ 
        
        Availability &  \num{15} & \num{1} & \num{2} & \num{3} & \num{2} & \num{3} \\ 
      
        Stability &  \num{11} & \num{3} & \num{1} & \num{2} & \num{0} & \num{2} \\ 
        
        Benchmarks &  \num{11} & \num{1} & \num{2} & \num{2} & \num{3} & \num{1} \\ 
        
        Ease of installation &  \num{10} & \num{1} & \num{1} & \num{5} & \num{0} & \num{2} \\ 
        
        Capability & \num{9} & \num{3} & \num{2} & \num{1} & \num{2} & \num{0} \\ 

        \hline
        \end{tabular}
\end{table}
\renewcommand{\arraystretch}{1}

We asked the participants to write about the limitations of their component selection process (Q13, N=\num{53/95}). The most common issue was that the process is \inlinecode{non-agile and time-consuming} (\num{28}). According to R99, \inlinequote{It is time consuming and sometimes we overlook some of the important features that we had to take care before selection.} Other complaints included \inlinecode{lack of access to reliable information} (\num{15}), \inlinecode{limited evaluation depth} (\num{10}), \inlinecode{reliance on human judgment} (\num{10}), \inlinecode{lack of assessing longevity} (\num{9}), and \inlinecode{lack of reliable automated tools} (\num{6}). 

We then asked the participants if they had used any tools in their software component evaluation process (Q14, N=\num{94/95}). The majority, \num{64}, reported that they did not use any tools, while \num{30} indicated that they had used some tools during evaluation. This is consistent with the literature, which shows that component selection tools are not in widespread use in industry.
\num{Twenty} participants who have used tools answered at least one of the \num{three} optional open-end sub-questions (Q15, N=\num{22/30}): \inlinequote{Which tools do you use and for which purpose,} \inlinequote{Which tools are AI-driven and what features do they offer,} and \inlinequote{Which tool have you found reliable and effective?} \inlinecode{ChatGPT} was mentioned \num{nine} times, making it the most frequently cited tool. It was used not only for component selection, as highlighted by R23---\inlinequote{ChatGPT can sometimes find a relevant library that we couldn't just find by searching}---but also for \inlinequote{generating test scenarios and codes} (R99) and \inlinequote{for quickly gaining a better understanding} (R8). Other tools included \inlinecode{SAP Digital Discovery Assessment}, \inlinecode{Snyk}, \inlinecode{SonarQube}, \inlinecode{Github Copilot}, and \inlinecode{Gemini}. 

We then asked: \inlinequote{Would you consider using AI-driven tools for software component evaluation in your organization?} (Q16, N=\num{95/95}). Most of the participants expressed a willingness to consider AI-driven tools (\num{64}), while \num{24} were unsure and \num{seven} participants stated that they would not consider using AI tools. People who were unwilling to use AI tools in component selection were asked just one further question (Q17a, N=\num{6/7}), namely their reason for opposing AI tool use. \num{Two} participants each specified \inlinename{insufficient understanding or experience in AI tools,} \inlinename{lack of trust in AI decisions,} and \inlinename{regulatory or compliance restrictions.} Other reasons included \inlinename{AI tools that do not meet their specific needs,} \inlinename{privacy concerns about data,} \inlinename{possible implementation costs,} and a \inlinename{preference for manual evaluation methods.} From their qualitative responses, we also identified \inlinecode{mistrust in the data bias,} \inlinecode{the lack of transparency of decision-making processes,} and the \inlinecode{reliability} of AI tools as additional concerns.
We used Python to conduct a Chi-Square test~\cite{franke2012chi} on participants' willingness to use AI tools and their level of familiarity with AI. The results showed a value of $\chi^2 = 21.25$ with a $p$-value of $6.5\times10^{-3}$, indicating a statistically significant relationship.
A Spearman correlation yielded $\rho = 0.41$ with a $p$-value of $3.94\times10^{-5}$, indicating a moderate positive relationship, i.e., people with more AI experience are somewhat more willing to use AI tools.
We also considered years of experience and country of residence, neither of which had any correlation with willingness to use AI tools in component selection.

\renewcommand{\arraystretch}{1} 

\begin{table}[htbp]
    \centering
    \caption{Severity of Concerns About AI Tools. \modified{Ordered by avg.}}
    \label{tab:importance_of_concerns}
    \begin{tabular}{L{0.57\linewidth}ccccc}

        \hline
        \textbf{Concerns} & 
        \multicolumn{5}{c}{\textbf{Importance}} \\ 
        \cline{2-6}
         & 
         \rotatebox{90}{\textbf{Extremely}} &
         \rotatebox{90}{\textbf{Very}} &
         \rotatebox{90}{\textbf{Moderately~}} & 
         \rotatebox{90}{\textbf{Slightly}} & 
         \rotatebox{90}{\textbf{Not at all}} \\

        \hline
        
        Concerns about data privacy & \num{23} & \num{14} & \num{9} & \num{4} & \num{1} \\ 
        Lack of trust in AI decisions & \num{16} & \num{13} & \num{15} & \num{5} & \num{0} \\ 
        Regulatory or compliance restrictions & \num{12} & \num{15} & \num{6} & \num{1} & \num{0} \\ 
        Concerns about data security & \num{19} & \num{6} & \num{4} & \num{1} & \num{0} \\ 
        Insufficient understanding or expertise in AI tools & \num{2} & \num{9} & \num{8} & \num{1} & \num{0} \\ 
        Potential cost of implementation & \num{4} & \num{9} & \num{4} & \num{0} & \num{0} \\ 
        Fear of job displacement & \num{3} & \num{5} & \num{7} & \num{2} & \num{0} \\ 
        Impact on existing workflows & \num{3} & \num{4} & \num{9} & \num{1} & \num{0} \\ 
        AI tools do not meet specific needs & \num{3} & \num{5} & \num{5} & \num{0} & \num{0} \\ 

        \hline
    \end{tabular}
\end{table}

\renewcommand{\arraystretch}{1} 
    
    \subsection{Perceptions of AI-driven Tools}

Next, we asked about the anticipated benefits of using AI-driven tools for the evaluation of software components (Q17b, N=\num{86/88}). Most of the participants (\num{70}) highlighted \inlinename{time savings} as the main advantage. \inlinename{Improve accuracy} (\num{57})  and \inlinename{predictive insights} (\num{41}) were also frequently mentioned. A smaller group (\num{7}) selected \inlinename{Others,} reflecting additional, less common expectations. However, \num{three} participants expected, to some extent, the \inlinecode{ability to find and predict vulnerabilities}, as noted by R27: \inlinequote{We are looking for vulnerabilities and security breaches.} These responses suggest that participants value AI tools primarily for their efficiency and ability to improve decision making accuracy.

Participants had several concerns about the integration of AI tools in their evaluation processes (Q18, N=\num{86/88}), such as \inlinename{data privacy} (\num{52}), \inlinename{lack of trust in AI decisions} (\num{50}), \inlinename{regulatory or compliance restrictions} (\num{34}), \inlinename{data security} (\num{30}), \inlinename{insufficient understanding of AI tools} (\num{21}), \inlinename{potential costs} (\num{18}), \inlinename{fears of job displacement} (\num{17}), \inlinename{impact on workflow} (\num{17}), and \inlinename{tools not meeting specific needs} (\num{13}). A smaller number of participants mentioned a \inlinename{preference for manual methods} (\num{9}), \inlinename{previous negative experiences} (\num{9}), and \inlinecode{lack of support for evaluating \modified{CBSD} models.} (\num{1}). The last answer was derived from QDA analysis. We then asked the participants to rank the severity of their concerns (Q19, N=\num{85/86}). \tblname{Table \ref{tab:importance_of_concerns}} shows the number of people who ranked each concern as \inlinequote{Extremely important} to \inlinequote{Not at all important.} Only top concerns with \num{50} or more points are shown, using a method of allocating \num{five} points for each \inlinequote{Extremely important} rating, \num{four} for each \inlinequote{Very important,} etc. 

While \inlinename{Concerns about data privacy} was the most pressing concern, with \num{23} respondents rating it \inlinequote{Extremely important} and \num{14} considering it \inlinequote{Very important,} other concerns---such as the \inlinename{potential cost of implementation,} \inlinename{AI tools that do not meet specific needs,} and \inlinename{fear of job displacement}---still appeared frequently in the \inlinequote{Very important} and \inlinequote{Moderately important} categories, underscoring the breadth of reservations about the integration of AI tools into software component selection processes.

    \subsection{Desired Features and Functionality}

In response to the question \inlinequote{Which features would you consider essential in an AI-driven tool for evaluating components,} participants selected several key features (Q20, N=\num{84/87}), including \inlinename{code quality analysis} (\num{54}), \inlinename{performance analytics} (\num{53}), \inlinename{security vulnerability scanning} (\num{43}), \inlinename{compatibility checks} (\num{41}), \inlinename{integration simulations} (\num{36}), \inlinename{ethical compliance indicators} (\num{21}), and \inlinename{community activity monitoring} (\num{20}). A smaller group provided their own arguments. \num{Two} participants expressed some  \inlinecode{hesitation about AI capabilities} in comparing components, while another \num{one} (R99) stated: \inlinequote{I need it predicts potential future problems based on the configuration.} 

We asked participants what features they would value in an AI tool (Q21, N=\num{83/87}). The most common selected ones were \inlinename{importance of recommendations} (\num{51}), \inlinename{need for alerts for potential issues} (\num{49}), \inlinename{Comparative analysis} between different components (\num{46}),  \inlinename{detailed reports} (\num{44}) and \inlinename{predictions or forecasts} (\num{39}).

\modified{We then asked} if they prefer an AI tool that integrates with their existing development environment (Q22, N=\num{79/87}). A significant majority, \num{64}, expressed a preference for integrated AI tools, and \num{15} participants preferred standalone tools. \num{Sixty-six} participants also shared their reasons. The main reasons for preferring an integrated AI tool were \inlinecode{enhancing productivity} (\num{22}), \inlinecode{increasing usability} (\num{21}), and \inlinecode{improving recommendations} (\num{13}). Meanwhile, the lack of trust in AI tools due to \inlinecode{intellectual property concerns} (\num{3}) was the most common reason among those with negative opinions. As R21 explained, \inlinequote{[my] company does not want AI to get sensitive data.}

    \subsection{Common Mistakes in Component Selection}

Participants highlighted several common mistakes they observed or experienced in their component selection processes (Q23, N=\num{78/87}). \tblname{Table \ref{tab:common_mistakes_by_stage}} lists the most common mistakes and how many people agreed that the mistake was common (italics column). \inlinename{Failing to assess long-term maintainability} was the most commonly identified mistake (\num{44}). Some participants provided their own insights, such as \inlinecode{choosing trendy components over well-suited ones} and \inlinecode{ignoring existing complaints on the internet}. Participants were asked to identify where in the software development process the component selection mistakes had the most impact (Q24, N=\num{70/77}). The remaining columns of \tblname{Table~\ref{tab:common_mistakes_by_stage}} shows the number of times a mistake was related to a particular stage of software development. For example, \inlinename{failure to assess long-term maintainability} was seen as a mistake occurring during \inlinequote{initial assessment of requirements} by \num{ten} participants.

\renewcommand{\arraystretch}{1.1} 
\addtolength{\tabcolsep}{-0.1em} 

\begin{table*}[htbp]
    \centering
    \caption{Common Mistakes as Identified by Participants and Categorized by Development Stages. \modified{Ordered by \# of votes.}}
    \label{tab:common_mistakes_by_stage}
    \begin{tabular}{lccccccccccccc}
    \hline
        \textbf{Mistake} & 
        \rotatebox{90}{\parbox{3cm}{\textbf{Number who voted it \\a common mistake}}}
        &
        
        \rotatebox{90}{\parbox{3cm}{\textbf{Initial assessment\\of requirements}}}
        & 
        \rotatebox{90}{\textbf{Finding components}} & 
        \rotatebox{90}{\parbox{3cm}{\textbf{Component\\selection}}} & 
        \rotatebox{90}{\parbox{3cm}{\textbf{Testing and\\validation}}} & 
        \rotatebox{90}{\parbox{3cm}{\textbf{Integration with\\other systems}}} & 
        \rotatebox{90}{\parbox{3cm}{\textbf{Performance\\evaluation}}} & 
        \rotatebox{90}{\textbf{Deployment}} & 
        \rotatebox{90}{\parbox{3cm}{\textbf{Maintenance\\and updates}}} & 
        \rotatebox{90}{\textbf{Documentation}} & 
        \rotatebox{90}{\parbox{3cm}{\textbf{Communication\\with stakeholders}}} & 
        \rotatebox{90}{\parbox{3cm}{\textbf{Project planning \\and scoping}}} & 
        \rotatebox{90}{\textbf{Others}} \\ 
        \hline
        
        Failure to assess long-term maintainability & \num{44} & \num{10} & \num{8} & \num{10} & \num{11} & \num{10} & \num{5} & \num{5} & \num{10} & \num{2} & \num{3} & \num{4} & \num{2} \\

        Inadequate evaluation of component compatibility & \num{30} & \num{6} & \num{9} & \num{12} & \num{12} & \num{6} & \num{2} & \num{4} & \num{1} & \num{0} & \num{0} & \num{3} & \num{0} \\
        
        Ignoring scalability requirements &  \num{29} & \num{7} & \num{3} & \num{4} & \num{4} & \num{2} & \num{4} & \num{3} & \num{5} & \num{1} & \num{3} & \num{4} & \num{0} \\

        Underestimating integration challenges & \num{25} & \num{3} & \num{5} & \num{8} & \num{8} & \num{11} & \num{3} & \num{4} & \num{1} & \num{3} & \num{1} & \num{2} & \num{0} \\

        Overlooking security vulnerabilities & \num{23} & \num{4} & \num{4} & \num{6} & \num{9} & \num{5} & \num{3} & \num{2} & \num{3} & \num{1} & \num{2} & \num{0} & \num{0} \\        
        
        Lack of thorough testing before integration & \num{22} & \num{1} & \num{2} & \num{5} & \num{13} & \num{3} & \num{2} & \num{1} & \num{2} & \num{0} & \num{0} & \num{0} & \num{0} \\
        
        Misalignment with project requirements or goals & \num{20} & \num{7} & \num{3} & \num{4} & \num{1} & \num{3} & \num{1} & \num{0} & \num{0} & \num{0} & \num{2} & \num{5} & \num{0} \\
        
        Overlooking licensing constraints & \num{18} & \num{2} & \num{2} & \num{3} & \num{2} & \num{2} & \num{0} & \num{1} & \num{0} & \num{1} & \num{1} & \num{1} & \num{1} \\

        Selecting components without proper vendor support & \num{18} & \num{1} & \num{3} & \num{3} & \num{1} & \num{2} & \num{0} & \num{2} & \num{2} & \num{2} & \num{0} & \num{1} & \num{0} \\
        
        \makecell[l]{Failing to involve key stakeholders in the\\decision-making process} & \num{17} & \num{5} & \num{3} & \num{5} & \num{1} & \num{2} & \num{1} & \num{1} & \num{1} & \num{1} & \num{2} & \num{1} & \num{1} \\

        Rushing the selection process due to time constraints & \num{16} & \num{5} & \num{7} & \num{4} & \num{3} & \num{4} & \num{2} & \num{0} & \num{2} & \num{1} & \num{2} & \num{1} & \num{0} \\

        Selecting based on cost alone & \num{15} & \num{7} & \num{5} & \num{5} & \num{1} & \num{2} & \num{2} & \num{1} & \num{1} & \num{1} & \num{1} & \num{1} & \num{0} \\
        
        Relying on insufficient or outdated documentation & \num{14} & \num{2} & \num{4} & \num{6} & \num{3} & \num{2} & \num{1} & \num{2} & \num{1} & \num{3} & \num{0} & \num{1} & \num{0} \\

        \hline
    \end{tabular}
\end{table*}

\renewcommand{\arraystretch}{1} 
\addtolength{\tabcolsep}{+0.1em} 

Participants described the consequences of incorrect selection of components in their projects (Q25, N=\num{35/73}). The most common impacts were \inlinecode{project delays or failure} (\num{24}), \inlinecode{wasted money and increased the costs} (\num{15}), the \inlinecode{need to reevaluate components} (\num{13}), \inlinecode{technical debt} (\num{10}), and \inlinecode{maintenance challenges} (\num{9}). In addition, \inlinecode{stakeholder dissatisfaction}, \inlinecode{trust issues in the product or development team}, and \inlinecode{missed business opportunities} were mentioned each \num{six} times. R76 listed \inlinequote{missing timelines, longer executive decision-making, and trust issues from management} as problems stemming from incorrect component selection.

Why do people select incorrect components (Q26, N=\num{71/73})? The most frequently cited reason was \inlinename{time pressure,} mentioned by \num{45} participants. \inlinename{Lack of information} (\num{40}), the \inlinename{complexity of the requirements} (\num{33}), \inlinename{lack of expertise} (\num{27}) and   \inlinename{poor communication among team members} (\num{22}) rounded out the top five reasons. \num{Four} participants provided other reasons, such as R23, who mentioned \inlinequote{employing very young developers who don't plan to stay on the project for more than a month or two.} 
 
In another open-ended question, participants were asked to describe the best practices, procedures, or tools they use to avoid common mistakes (Q27, N=\num{31/73}). The most frequently mentioned strategies were \inlinecode{thorough assessment in advance} (\num{23}), \inlinecode{involving stakeholders} (\num{15}), \inlinecode{relying on reliable sources of information} (\num{14}), and \inlinecode{consulting experts} (\num{11}). According to R14, \inlinequote{to avoid common mistakes, I ensure clear requirements, maintain regular communication with stakeholders, and implement iterative testing to catch issues early.} Some best practices such as \inlinecode{collaboration with AI} (\num{46}) were explicitly mentioned. R77 wrote, \inlinequote{I prefer to talk to senior developers or ask AI.}

We asked the participants if they believe AI can help identify and mitigate some of the common mistakes in component selection (Q28, N=\num{74/81}). Most of the participants expressed optimism, with \num{44} responding \inlinequote{yes.} \num{Twenty-eight} participants were more cautious, responding  \inlinequote{maybe,} suggesting that they see potential but may have hesitations. \num{Two} participants answered \inlinequote{no.} We asked for a vote of confidence on AI's ability to help mitigate common mistakes in component selection (Q29, N=\num{70/74}). \tblname{Table \ref{tab:ai_help_mitigate_mistakes}} lists the mistakes, along with how many people \inlinequote{strongly agreed \modified{(++)},} \inlinequote{somewhat agreed \modified{(+)},} \inlinequote{were neutral \modified{(0)},} \inlinequote{somewhat disagreed \modified{(-)},} or \inlinequote{strongly disagreed \modified{(-{}-)}} that AI had the potential to remediate the mistake.

    \subsection{Role of AI in Mitigating Mistakes}

We asked participants about their experiences with hallucinations in AI tools (Q30, N=\num{76/80}). Hallucination occurs AI tools provide incorrect or misleading information, such as recommending a component which doesn't exist. Most (\num{58}) participants reported encountering hallucinations, while \num{18}  stated that they had not experienced this problem. The most common approach for addressing AI hallucinations (Q31, N=\num{41/79}) was \inlinecode{manual validation and correction} (\num{38}), followed by \inlinecode{cross-referencing with traditional sources} such as official documentation, forums, StackOverflow or search engines (\num{22}), \inlinecode{cross-verification with another AI tool} (\num{5}), and \inlinecode{using question engineering techniques} (\num{19}), which included \inlinecode{adaptive prompting} (\num{13}), \inlinecode{question refinement} (\num{5}), and \inlinecode{testing across input variations} (\num{1}). R55 described \inlinecode{testing across input variations}: \inlinequote{[I] test AI with different formats of requirements to see if answers are same.} Other strategies people used to validate AI responses (Q32, N=\num{36/79}) included \inlinecode{using a test-driven approach} (\num{10}) and \inlinecode{consulting with senior experts} (\num{5}) R53 explained their company's human validation approach: \inlinequote{I [was] employ[ed] to verify the accuracy of AI-generated recommendations~\ldots}

    \subsection{Challenges with AI tools}

Participants were asked to describe their best practices for using AI tools in their component selection workflow (Q33, N=\num{36/79}). In general, there appear to be no widely recognized guidelines or best practices for using AI tools among practitioners, since \num{ten} respondents stated that they had \inlinecode{no best practices}. \num{Twelve} recommended an \inlinecode{interactive engagement with AI} which means using AI tools in the workflow iteratively. Many repeated their existing validation processes as best practices, such as \inlinecode{manual check}, \inlinecode{cross-referencing}, and \inlinecode{testing with a separate small project}, reflecting a degree of \inlinecode{skepticism toward AI recommendations}. For example, R55 gave \inlinequote{Evaluation with human findings} as their best practice.

We were also interested in industry expectations for AI tools in component selection and asked participants how such tools could be designed (Q34, N=\num{34/79}). The themes most commonly mentioned were \inlinecode{providing a better user experience} (\num{30}), \inlinecode{improving productivity} (\num{29}), \inlinecode{ensuring accurate and reliable results} (\num{21}), \inlinecode{offering transparency and justification} (\num{16}), \inlinecode{allowing human oversight} (\num{12}), and \inlinecode{providing automation} (\num{11}). R4 recommended that AI tools must \inlinequote{guide the user to communicate with the AI tool efficiently by following a step-by-step Q\&A approach.} Other themes  included \inlinecode{compatibility with existing workflows} (\num{9}), \inlinecode{leaving the final decision for humans} (\num{8}), \inlinecode{the ability to predict errors} (\num{7}), \inlinecode{having acceptable performance} (\num{5}),  \inlinecode{security and vulnerability detection and prediction} (\num{4}), \inlinecode{better natural language understanding} (\num{3}), \inlinecode{respect for intellectual property} (\num{3}), \inlinecode{outperforming search engines} (\num{2}),  \inlinecode{affordability} (\num{2}).  R40 suggested comprehension of versioning, stating, \inlinequote{AI should give [a] solution, that works in [an] environment, that was updated last time 5 years ago.} Another expectation was regular updates, as described by R44: \inlinequote{\ldots continuously refine AI algorithms to adapt to evolving requirements and reduce biases.}

    \subsection{Development Considerations}

Participants shared their views on the most valuable data sources for training AI models in relation to component selection (Q35, N=\num{74/79}). \inlinename{Code repositories} were the main choice, highlighted by \num{55} participants, followed by \inlinename{issue trackers,} which were considered valuable by \num{42}. \inlinename{User reviews} were also a significant source, selected by \num{39} participants, while \inlinename{forums} were mentioned by \num{22}. Relying on \inlinecode{internal forums, communities or databases} were noted \num{four} times in response to this question.

The participants shared how willing they are to share project data for the development of AI tools (Q36, N=\num{74/79}). The results of this Likert-scale question, in order of least to most favorable, were \inlinename{not willing at all} (\num{21}), \inlinename{slightly willing} (\num{18}), \inlinename{moderately willing} (\num{16}), \inlinename{very willing} (\num{15}), and \inlinename{extremely willing} (\num{4}). 

The most important criteria for validating AI-driven tools for component selection (Q37, N=\num{74/79}) were \inlinename{accuracy} (\num{46}) and \inlinename{reliability} (\num{46}).  \inlinename{Ease of use} was also highly valued (\num{37}). \inlinename{Integration capabilities,} reflecting the need for seamless adoption within existing workflows, were noted by \num{23} participants. The expectations mentioned in response to Q34 can also be considered criteria for evaluating AI component selection tools.

\begin{table}[htbp]
    \centering
    \caption{AI's Potential to Mitigate Mistakes. \modified{Ordered by weighted scores (e.g., for the first row: 2×5+1×19-1×5-2×1=22).}}
    \label{tab:ai_help_mitigate_mistakes}
    \begin{tabular}{L{0.61\linewidth}ccccc}
    
       \hline
        \textbf{Mistake} & 
        \multicolumn{5}{c}{\textbf{Agreement}} \\
        \cline{2-6}
         &  
        \textbf{++} &
        \textbf{+} &
        \textbf{0} &
        \textbf{-} &
        \textbf{-{}-} \\
        \hline

        Failure to assess long-term maintainability & \num{5} & \num{19} & \num{8} & \num{5} & \num{1} \\ 
        Rushing the selection process due to time constraints & \num{5} & \num{9} & \num{0} & \num{1} & \num{0} \\ 
        Overlooking security vulnerabilities & \num{5} & \num{9} & \num{3} & \num{2} & \num{0} \\ 
        Inadequate evaluation of component compatibility & \num{4} & \num{12} & \num{7} & \num{2} & \num{1} \\ 
        Misalignment with project requirements or goals & \num{3} & \num{10} & \num{1} & \num{2} & \num{0} \\ 
        Lack of thorough testing before integration & \num{3} & \num{8} & \num{4} & \num{2} & \num{1} \\ 
        Relying on insufficient or outdated documentation & \num{1} & \num{8} & \num{2} & \num{0} & \num{1} \\ 
        Overlooking licensing constraints & \num{3} & \num{5} & \num{4} & \num{2} & \num{1} \\ 
        Underestimating integration challenges & \num{3} & \num{9} & \num{2} & \num{5} & \num{2} \\ 
        Ignoring scalability requirements & \num{0} & \num{12} & \num{6} & \num{3} & \num{2} \\ 
        Selecting based on cost alone & \num{2} & \num{5} & \num{5} & \num{0} & \num{2} \\ 
        Selecting components without proper vendor support & \num{2} & \num{4} & \num{3} & \num{2} & \num{2} \\ 

    \hline
    \end{tabular}
\end{table}

    \subsection{Adoption and Integration}

How important it is for an AI tool to provide clear explanations for its recommendations, making the decision-making process understandable to humans (Q38, N=\num{74/79})? Most of the participants valued \inlinequote{explainability,} with \num{31} rating it as \inlinename{very important} and \num{19} considering it \inlinename{extremely important.} \num{14} participants found it \inlinename{moderately important,} \num{eight} found it \inlinename{slightly important} and \num{two} said it was \inlinename{not important.} 

Participants shared their thoughts on the support or training they would need to effectively use an AI-driven tool for component selection (Q39, N=\num{71/79}). In order of preference, respondants indicated an interest in \inlinename{clear examples of best practices} (\num{41}), \inlinename{interactive tutorials or video guides} (\num{31}), \inlinename{detailed user manuals or documentation} (\num{26}), \inlinename{hands-on training sessions or workshops} (\num{24}), \inlinename{training tailored to specific workflows} (\num{19}), \inlinename{customer support or a help-desk} (\num{17}), \inlinename{on-boarding assistance} (\num{11}), and \inlinename{regular webinars or live questions and answers (Q\&A) sessions} (\num{11}). Interestingly, \num{11} participants felt that \inlinename{no additional support or training} was necessary.

Finally, we asked about concerns about the long-term effects of AI-driven tools (Q40, N=\num{73/79}). The most common concern, mentioned by \num{37} participants, was a \inlinename{lack of transparency or fairness in AI decision-making.} Other concerns included \inlinename{over-reliance on AI,} which could lead to skill atrophy (\num{32}), concerns about \inlinename{job displacement or reduced need for human expertise} (\num{30}), \inlinename{security and privacy risks related to data usage} (\num{30}), \inlinename{potential misuse or unethical applications of AI} (\num{23}), \inlinename{limited adaptability to unforeseen scenarios} (\num{22}), the \inlinename{environmental impact of AI computing resources} (\num{18}), and \inlinename{difficulty of keeping up with the rapid technological changes} (\num{16}). \num{Thirteen} participants reported \inlinename{no concerns at this time}. \num{One} participant, R29, shared a unique concern, which was \inlinequote{lack of accountability for decisions.}

    \section{Discussion}\label{sec:discussion}
        Our survey illuminated several insights about component selection, particularly regarding the use of AI tools in the selection process of software components. 

In answer to \textbf{RQ1}, \inlinequote{What are the most frequently used methods for component selection in practice,} we found that the most common component selection method is ad hoc decision making\modified{, while also revealing an emerging tendency toward AI-driven approaches}. Only \modified{two academia participants} mentioned specific algorithmic techniques, namely Fuzzy ANP and MCDM. This aligns with previous observations that formal methods are rarely used in industry \cite{ayala2011selection, petersen2017choosing}. However, some practitioners are employing systematic techniques, including sequential evaluation, market analysis, using proof-of-concept exploration or test-driven approaches, and utilizing demonstrations, which suggests that there is some interest among developers in employing structured approaches.

\textbf{RQ2} was \inlinequote{What tools do practitioners rely on for component selection?} None of our participants identified a specific tool, which also mirrors previous findings \cite{mahapatro2024reviewing}. People did mention using tools in an auxiliary capacity, for instance in OSS component evaluation, where Snyk and SonarQube were highlighted. At the time of this survey (before the popularity of DeepSeek), ChatGPT was the most frequently reported AI tool for searching the internet to find alternatives, generating test cases, and evaluating documentations and component information. Other less popular platforms which participants found useful were  Hugging Face, Deepeval, Tabnine, DeepCode, CodeAI, ClientAI, and CursorAI. However, concerns about the validity of the results and hallucinations of AI were reported, which required manual verification against reliable sources such as official documentation or expert consultations. Some respondents also used AI-enabled verification techniques, such as cross-checking results from two different AI tools (e.g., Gemini and ChatGPT) or assigning different tasks to AI models, such as using ChatGPT for information gathering and GitHub Copilot for generating unit tests. None of these approaches is fully automated, and they all still require significant manual effort to ensure accuracy and reliability. This led \num{11} respondents to specifically call for automated component selection tools. This agrees with previous studies that reported that industry is looking for validated automated state-of-the-art decision making tools~\cite{mahapatro2024reviewing, farshidi2018decision, badampudi2018decision, siena2014modelling}.
In addition, \num{81\%} of the respondents expressed their interest in integrated AI-driven component selection tools in IDEs. This shows that there is a growing interest in using AI in component selection.

One of the gaps in \modified{CBSD} studies relates to quality criteria for comparing components; there is no complete and widely accepted list of criteria for component selection nor solid advice on the prioritization of quality attributes~\cite{vale2016twenty,nabot2024software,badampudi2016software}. Given the vast number of requirements and the dynamic nature of software engineering, the importance of certain criteria may change over time. Some may become less relevant or even obsolete, while new ones may emerge based on new technologies, such as defining standards for assessing ML and AI components~\cite{ali2022systematic}. \textbf{RQ3} aimed to address this concern by asking \inlinequote{What quality criteria are prioritized in the industry for the selection of components?} Based on the number of votes, the most important criteria were reliability, active development or community activity, quality of documentation, security, cost, and ease of use. However, when considering prioritization in a sequential evaluation process, the top ranked criteria were active development or community activity, security, reliability, cost, and architectural considerations. 

The emphasis on \inlinename{active development or community activity} suggests that the industry focuses primarily on OSS components since  development is not easily measured for closed source components, nor is it as meaningful a metric. The emphasis on  documentation quality indicates that practitioners prefer to find answers directly in supporting materials rather than relying on community discussions or help desks, which may involve delays. Although evaluating community activity reflects the importance of community support, the high value placed on documentation quality highlights the need for independence when resolving issues. Furthermore, the overlap between both prioritized lists reinforces that reliability, security, and cost remain strong concerns in component selection. This is aligned with the results of other studies~\cite{chatzipetrou2018component,borg2019selecting, badampudi2018decision}, but at the same time the variation in responses shows that requirements are not universal and any solution must take into consideration the user's distinct needs.

In response to \textbf{RQ4}, \inlinequote{What best practices are followed in the industry for the successful selection of components,} the survey results revealed that best practices include a thorough evaluation of the components, involving stakeholders and collecting their feedback, relying on reliable sources of information, and consulting experts. Although traditional evaluation methods remain dominant, some participants explicitly mentioned collaboration with AI as an emerging best practice. However, the way AI is used in component selection reflects a cautious approach rather than full automation. Participants shared techniques to mitigate AI limitations, such as verifying AI-generated recommendations by cross-verification with another AI tool or cross-referencing results with traditional sources such as search engines and documentation.

Generative AI has led to significant upheaval, both within and outside of the software industry. We saw that while some practitioners expressed concerns about job loss and degradation of skills, others were embracing it as part of the component selection process, with one person even reporting that they had been hired to verify AI output. \citeauthor{mahapatro2024reviewing} (\citeyear{mahapatro2024reviewing}) advocated for the development of AI tools to assist CBSE~\cite{mahapatro2024reviewing}, and following their lead we posed \textbf{RQ5}, asking: \inlinequote{How do professionals perceive AI-driven tools for component selection, particularly in terms of barriers, enablers, adoption, integration, and intended features?} Practitioners expect AI tools to be designed to improve productivity by providing a better user experience, ensure accurate and reliable results, offer transparency and justification, allow human oversight, and enable automation. They must be affordable, have acceptable performance, and outperform search engines in finding relevant information.

The features that mattered the most to respondents were seamless integration into existing workflows, the ability to predict errors and detect security vulnerabilities, powerful natural language processing capabilities, and clear guidelines for respecting intellectual property. These expectations reflect a cautious but strategic approach to AI adoption, where practitioners see AI as a tool to enhance efficiency rather than replace human expertise. The emphasis on transparency, human oversight, and intellectual property protection highlights the industry's concerns about trust, accountability, and legal risks in AI-driven decision-making.

\modified{Researchers can build on these findings to refine methods that fit current processes, while practitioners can adopt cross-verification and clear oversight policies. These steps may help reduce risks and make AI-driven solutions more effective in real-world settings.}

    \section{Limitations}\label{sec:limitations}        
        To evaluate the limitations of our work, we used \citeauthor{wohlin2012experimentation}'s framework \citeyear{wohlin2012experimentation}~\cite{wohlin2012experimentation}. 
 
\subsection{Construct validity} 

\subsubsection{Survey Length and Question Design} 
Our survey showed up to \num{40} questions, which required \num{20–30} minutes for the most complete path. Feedback indicated that the survey felt long; some participants dropped out or skipped sections due to time constraints. Although branching questions reduced the burden for certain paths, the overall length may have influenced how carefully the participants answered. \modified{Furthermore, we did not include explicit attention checks or similar features to control the quality of responses, introducing a risk of unreliable answers. However, the option to leave questions unanswered likely reduced the impact of inattentive participation.}

\subsubsection{Question Interpretation}
To capture various perspectives, we included multiple choice, single choice, matrix, open-ended, and ordering questions. We observed a dropout between Q10 (selecting criteria) with \num{98} records and Q11 (ordering selected criteria) with \num{78} records. We believe that the dropout was because some people (\num{20}) were not used to ordering questions and because the questions were optional, they left the question unanswered. 

\subsection{Internal Validity}

\subsubsection{Dropout Rate and Missing Data}
Of the \num{99} participants who initially consented, \num{85} \modified{reached the last page of the survey} (\num{15\%} dropout). It could bias the findings if only more motivated respondents completed all the items. We conducted statistical tests to examine whether years of experience or geographic location influenced survey dropout patterns. Participants who dropped out were not significantly different in terms of experience level \modified{(Mann-Whitney U-test, $p = 0.86$)} or country \modified{(Chi-square, $\chi^2 = 16.72$, $p = 0.40$)} from those who completed the survey. Furthermore, there was no clear pattern showing that participants with less experience \modified{(Spearman, $\rho = -0.13$, $p = 0.65$)} or from a specific country \modified{(Kruskal-Wallis, $H = 5.25$, $p = 0.63$)} dropped out earlier than those with more experience or located in other countries.

\modified{
We also compared the responses of participants from academia and industry across \num{25} single- and multiple-choice questions using Chi-square tests and found no meaningful differences in \num{24} of them.

The only exception was about the importance of explainable recommendations for AI tools
(Q38, $\chi^{2} = \num{9.78}$, $p = \num{0.04}$).
Most researchers and half of non-researchers selected \inlinecode{Very important}, while the remaining non-researchers selected \inlinecode{Extremely} or \inlinecode{Moderately important}. The wider variability (but similar average) seems unlikely to affect overall results.
}

\subsection{External Validity}

\subsubsection{Sampling and Recruitment}
We sent approximately \num{500} targeted invitations. We cannot calculate our response rate because we cannot confirm how many recipients received or opened the invitation emails, or how many were contacted through snowballing recruitment or saw LinkedIn posts. Convenience sampling can limit the diversity of our sample, as participants self-selected based on their interest and availability. We attempted to expand our sample through multiple forms of recruitment to reduce this concern. \modified{We also had no control mechanism to ensure that all participants met the inclusion and exclusion criteria outlined in Section \ref{sub-sub-sec-sampling-approach}.}

\subsubsection{Generalization Across Contexts}
Although the number of participants (\num{85} completed) is reasonable according to the standards of SE surveys\modified{~\cite{baltes2022sampling}}, the study may not cover all the communities of practitioners or geographic regions. Cultural and organizational differences can affect how developers approach component selection, and our sample may not reflect the entire global developer population. We sought to mitigate this concern by collecting not only quantitative data but also qualitative data, which can help reveal a wider breadth of perspective.

\subsection{Conclusion Validity}

\subsubsection{Coding Process and Interpretation}
\num{Two} coders independently analyzed a portion of the qualitative responses, applying an inter-rater agreement threshold of \num{80\%} to ensure consistent coding. 
The majority of the data was only coded by the first author, although we practiced inter-rater reliability by involving the second author throughout the process after achieving respectable inter-rater agreement in an early iteration. Survey data is also highly structured, which provides some measure of certainty concerning the consistency of the analysis.

\subsubsection{Data Analysis Methods}
We used descriptive statistics for quantitative data and thematic analysis for qualitative data. Although these methods are suitable for exploratory survey research, any fault in data preparation, coding, or interpretation could be a threat to the validity of our conclusions. In order allow for a thorough evaluation of our approach, we have provided the data as part of our replication package.

\section{Conclusion}\label{sec:conclusion}
    This study explored how professionals select software components, the tools they use, and their expectations for AI in this process. The findings show that ad hoc methods remain the most common approach, with manual validation and expert consultation being key best practices. Although AI tools such as ChatGPT are increasingly being used for searching, testing, and documentation review, concerns about accuracy, reliability, and hallucinations highlight the need for manual verification.

Practitioners expect AI tools to improve productivity, ensure transparency, provide reliable results, and integrate seamlessly into existing workflows. Security, cost, active development, and documentation quality were identified as the most critical criteria in component selection. In general, while AI has potential in this field, fully automated solutions are still lacking and human oversight remains essential. Future work should focus on developing reliable, efficient, and explainable AI-driven tools for component selection, while continuing to take industry perspectives into consideration.

Future research could also explore component selection practices in particular regions or industries to determine if there are significant differences to the more general results we obtained. Repeating the study in a couple of years could also track how AI adoption in this field evolves over time as tools improve.

This study provides several contributions to understanding component selection and AI adoption in CBSD.

\begin{itemize}
    \item \textbf{Capturing the industry perspective:} Unlike many previous studies that focus on technical methods without industry involvement, our research directly captures the practitioner's perspective, providing real-world insights into how developers evaluated, selected, and verified software components in late 2024. This ensures that our findings are grounded in actual industry challenges.
    \item \textbf{Bridging the tool support gap for component selection:} We identified the need to establish better practical, transparent and simple formal component selection methods that can be implemented by the industry. In particular, automated selection tools can fill this gap. 
    \item \textbf{Identifying key quality criteria for component selection:} Our results show how the trend toward utilizing OSS components and developer independence has changed the prioritization of certain quality criteria, such as documentation quality, active development, and community activity, so that they are as important as security, reliability, and cost.
    \item \textbf{Clarifying the role of AI and industry expectations for future tools:} Our findings show that AI is currently seen as an auxiliary tool instead of a decision maker. Most practitioners cross-verify AI-generated output rather than fully relying on them. Practitioners expect AI tools to be transparent, accurate, and well-integrated into existing workflows. However, respecting intellectual property and affordability play a role in acceptance of such tools.
\end{itemize}
    
\section*{Acknowledgments}
    \modified{We would like to thank Musa Taib, Alireza Amini, Nabi Zamani, Alireza Esmaili, and Amr Qura for promoting the survey and all the participants for answering.}

\bibliographystyle{ACM-Reference-Format}
\bibliography{main}


\begin{thebibliography}{38}


\ifx \showCODEN    \undefined \def \showCODEN     #1{\unskip}     \fi
\ifx \showDOI      \undefined \def \showDOI       #1{#1}\fi
\ifx \showISBNx    \undefined \def \showISBNx     #1{\unskip}     \fi
\ifx \showISBNxiii \undefined \def \showISBNxiii  #1{\unskip}     \fi
\ifx \showISSN     \undefined \def \showISSN      #1{\unskip}     \fi
\ifx \showLCCN     \undefined \def \showLCCN      #1{\unskip}     \fi
\ifx \shownote     \undefined \def \shownote      #1{#1}          \fi
\ifx \showarticletitle \undefined \def \showarticletitle #1{#1}   \fi
\ifx \showURL      \undefined \def \showURL       {\relax}        \fi
\providecommand\bibfield[2]{#2}
\providecommand\bibinfo[2]{#2}
\providecommand\natexlab[1]{#1}
\providecommand\showeprint[2][]{arXiv:#2}

\bibitem[Ali et~al\mbox{.}(2022)]%
        {ali2022systematic}
\bibfield{author}{\bibinfo{person}{Mohamed~Abdullahi Ali}, \bibinfo{person}{Ng~Keng Yap}, \bibinfo{person}{Abdul Azim~Abd Ghani}, \bibinfo{person}{Hazura Zulzalil}, \bibinfo{person}{Novia~Indriaty Admodisastro}, {and} \bibinfo{person}{Amin~Arab Najafabadi}.} \bibinfo{year}{2022}\natexlab{}.
\newblock \showarticletitle{A systematic mapping of quality models for AI systems, software and components}.
\newblock \bibinfo{journal}{\emph{Applied Sciences}} \bibinfo{volume}{12}, \bibinfo{number}{17} (\bibinfo{year}{2022}), \bibinfo{pages}{8700}.
\newblock


\bibitem[Ayala et~al\mbox{.}(2011)]%
        {ayala2011selection}
\bibfield{author}{\bibinfo{person}{Claudia Ayala}, \bibinfo{person}{{\O}yvind Hauge}, \bibinfo{person}{Reidar Conradi}, \bibinfo{person}{Xavier Franch}, {and} \bibinfo{person}{Jingyue Li}.} \bibinfo{year}{2011}\natexlab{}.
\newblock \showarticletitle{Selection of third party software in Off-The-Shelf-based software development—An interview study with industrial practitioners}.
\newblock \bibinfo{journal}{\emph{Journal of Systems and Software}} \bibinfo{volume}{84}, \bibinfo{number}{4} (\bibinfo{year}{2011}), \bibinfo{pages}{620--637}.
\newblock


\bibitem[Badampudi et~al\mbox{.}(2018)]%
        {badampudi2018decision}
\bibfield{author}{\bibinfo{person}{Deepika Badampudi}, \bibinfo{person}{Krzysztof Wnuk}, \bibinfo{person}{Claes Wohlin}, \bibinfo{person}{Ulrik Franke}, \bibinfo{person}{Darja Smite}, {and} \bibinfo{person}{Antonio Cicchetti}.} \bibinfo{year}{2018}\natexlab{}.
\newblock \showarticletitle{A decision-making process-line for selection of software asset origins and components}.
\newblock \bibinfo{journal}{\emph{Journal of Systems and Software}}  \bibinfo{volume}{135} (\bibinfo{year}{2018}), \bibinfo{pages}{88--104}.
\newblock


\bibitem[Badampudi et~al\mbox{.}(2016)]%
        {badampudi2016software}
\bibfield{author}{\bibinfo{person}{Deepika Badampudi}, \bibinfo{person}{Claes Wohlin}, {and} \bibinfo{person}{Kai Petersen}.} \bibinfo{year}{2016}\natexlab{}.
\newblock \showarticletitle{Software component decision-making: In-house, OSS, COTS or outsourcing-A systematic literature review}.
\newblock \bibinfo{journal}{\emph{Journal of Systems and Software}}  \bibinfo{volume}{121} (\bibinfo{year}{2016}), \bibinfo{pages}{105--124}.
\newblock


\bibitem[Baltes and Ralph(2022)]%
        {baltes2022sampling}
\bibfield{author}{\bibinfo{person}{Sebastian Baltes} {and} \bibinfo{person}{Paul Ralph}.} \bibinfo{year}{2022}\natexlab{}.
\newblock \showarticletitle{Sampling in software engineering research: A critical review and guidelines}.
\newblock \bibinfo{journal}{\emph{Empirical Software Engineering}} \bibinfo{volume}{27}, \bibinfo{number}{4} (\bibinfo{year}{2022}), \bibinfo{pages}{94}.
\newblock


\bibitem[Borg et~al\mbox{.}(2019)]%
        {borg2019selecting}
\bibfield{author}{\bibinfo{person}{Markus Borg}, \bibinfo{person}{Panagiota Chatzipetrou}, \bibinfo{person}{Krzysztof Wnuk}, \bibinfo{person}{Emil Al{\'e}groth}, \bibinfo{person}{Tony Gorschek}, \bibinfo{person}{Efi Papatheocharous}, \bibinfo{person}{Syed Muhammad~Ali Shah}, {and} \bibinfo{person}{Jakob Axelsson}.} \bibinfo{year}{2019}\natexlab{}.
\newblock \showarticletitle{Selecting component sourcing options: a survey of software engineering’s broader make-or-buy decisions}.
\newblock \bibinfo{journal}{\emph{Information and Software Technology}}  \bibinfo{volume}{112} (\bibinfo{year}{2019}), \bibinfo{pages}{18--34}.
\newblock


\bibitem[Braun and Clarke(2006)]%
        {braun2006using}
\bibfield{author}{\bibinfo{person}{Virginia Braun} {and} \bibinfo{person}{Victoria Clarke}.} \bibinfo{year}{2006}\natexlab{}.
\newblock \showarticletitle{Using thematic analysis in psychology}.
\newblock \bibinfo{journal}{\emph{Qualitative research in psychology}} \bibinfo{volume}{3}, \bibinfo{number}{2} (\bibinfo{year}{2006}), \bibinfo{pages}{77--101}.
\newblock


\bibitem[Chatzipetrou et~al\mbox{.}(2018)]%
        {chatzipetrou2018component}
\bibfield{author}{\bibinfo{person}{Panagiota Chatzipetrou}, \bibinfo{person}{Emil Al{\'e}groth}, \bibinfo{person}{Efi Papatheocharous}, \bibinfo{person}{Markus Borg}, \bibinfo{person}{Tony Gorschek}, {and} \bibinfo{person}{Krzysztof Wnuk}.} \bibinfo{year}{2018}\natexlab{}.
\newblock \showarticletitle{Component selection in Software Engineering-Which attributes are the most important in the decision process?}. In \bibinfo{booktitle}{\emph{2018 44th Euromicro Conference on Software Engineering and Advanced Applications (SEAA)}}. \bibinfo{publisher}{IEEE}, \bibinfo{address}{Prague, Czech Republic}, \bibinfo{pages}{198--205}.
\newblock


\bibitem[Chatzipetrou et~al\mbox{.}(2020)]%
        {chatzipetrou2020component}
\bibfield{author}{\bibinfo{person}{Panagiota Chatzipetrou}, \bibinfo{person}{Efi Papatheocharous}, \bibinfo{person}{Krzysztof Wnuk}, \bibinfo{person}{Markus Borg}, \bibinfo{person}{Emil Al{\'e}groth}, {and} \bibinfo{person}{Tony Gorschek}.} \bibinfo{year}{2020}\natexlab{}.
\newblock \showarticletitle{Component attributes and their importance in decisions and component selection}.
\newblock \bibinfo{journal}{\emph{Software quality journal}}  \bibinfo{volume}{28} (\bibinfo{year}{2020}), \bibinfo{pages}{567--593}.
\newblock


\bibitem[Chen et~al\mbox{.}(2007)]%
        {chen2007survey}
\bibfield{author}{\bibinfo{person}{Weibing Chen}, \bibinfo{person}{Jingyue Li}, \bibinfo{person}{Jianqiang Ma}, \bibinfo{person}{Reidar Conradi}, \bibinfo{person}{Junzhong Ji}, {and} \bibinfo{person}{Chunnian Liu}.} \bibinfo{year}{2007}\natexlab{}.
\newblock \showarticletitle{A survey of software development with open source components in Chinese software industry}. In \bibinfo{booktitle}{\emph{Software Process Dynamics and Agility: International Conference on Software Process, ICSP 2007, Minneapolis, MN, USA, May 19-20, 2007. Proceedings}}. \bibinfo{publisher}{Springer}, \bibinfo{address}{Minneapolis, MN, USA}, \bibinfo{pages}{208--220}.
\newblock


\bibitem[Cohen et~al\mbox{.}(1960)]%
        {cohen:1960:coefficient}
\bibfield{author}{\bibinfo{person}{Jacob Cohen} {et~al\mbox{.}}} \bibinfo{year}{1960}\natexlab{}.
\newblock \showarticletitle{A coefficient of agreement for nominal scales}.
\newblock \bibinfo{journal}{\emph{Educational and psychological measurement}} \bibinfo{volume}{20}, \bibinfo{number}{1} (\bibinfo{year}{1960}), \bibinfo{pages}{37--46}.
\newblock


\bibitem[Cruzes and Dyba(2011)]%
        {cruzes2011recommended}
\bibfield{author}{\bibinfo{person}{Daniela~S Cruzes} {and} \bibinfo{person}{Tore Dyba}.} \bibinfo{year}{2011}\natexlab{}.
\newblock \showarticletitle{Recommended steps for thematic synthesis in software engineering}. In \bibinfo{booktitle}{\emph{2011 international symposium on empirical software engineering and measurement}}. \bibinfo{publisher}{IEEE}, \bibinfo{address}{Banff, AB, Canada}, \bibinfo{pages}{275--284}.
\newblock


\bibitem[Farshidi et~al\mbox{.}(2018)]%
        {farshidi2018decision}
\bibfield{author}{\bibinfo{person}{Siamak Farshidi}, \bibinfo{person}{Slinger Jansen}, \bibinfo{person}{Rolf De~Jong}, {and} \bibinfo{person}{Sjaak Brinkkemper}.} \bibinfo{year}{2018}\natexlab{}.
\newblock \showarticletitle{A decision support system for cloud service provider selection problem in software producing organizations}. In \bibinfo{booktitle}{\emph{2018 IEEE 20th Conference on Business Informatics (CBI)}}, Vol.~\bibinfo{volume}{1}. \bibinfo{publisher}{IEEE}, \bibinfo{address}{Vienna, Austria}, \bibinfo{pages}{139--148}.
\newblock


\bibitem[Fatima et~al\mbox{.}(2017)]%
        {fatima2017risk}
\bibfield{author}{\bibinfo{person}{Fariha Fatima}, \bibinfo{person}{Saqib Ali}, {and} \bibinfo{person}{Muhammad~Usman Ashraf}.} \bibinfo{year}{2017}\natexlab{}.
\newblock \showarticletitle{Risk Reduction Activities Identification in Software Component Integration for Component Based Software Development (CBSD)}.
\newblock \bibinfo{journal}{\emph{International Journal of Modern Education and Computer Science}} \bibinfo{volume}{9}, \bibinfo{number}{4} (\bibinfo{year}{2017}), \bibinfo{pages}{19}.
\newblock


\bibitem[Franke et~al\mbox{.}(2012)]%
        {franke2012chi}
\bibfield{author}{\bibinfo{person}{Todd~Michael Franke}, \bibinfo{person}{Timothy Ho}, {and} \bibinfo{person}{Christina~A Christie}.} \bibinfo{year}{2012}\natexlab{}.
\newblock \showarticletitle{The chi-square test: Often used and more often misinterpreted}.
\newblock \bibinfo{journal}{\emph{American journal of evaluation}} \bibinfo{volume}{33}, \bibinfo{number}{3} (\bibinfo{year}{2012}), \bibinfo{pages}{448--458}.
\newblock


\bibitem[Harutyunyan et~al\mbox{.}(2019)]%
        {harutyunyan2019industry}
\bibfield{author}{\bibinfo{person}{Nikolay Harutyunyan}, \bibinfo{person}{Andreas Bauer}, {and} \bibinfo{person}{Dirk Riehle}.} \bibinfo{year}{2019}\natexlab{}.
\newblock \showarticletitle{Industry requirements for FLOSS governance tools to facilitate the use of open source software in commercial products}.
\newblock \bibinfo{journal}{\emph{Journal of Systems and Software}}  \bibinfo{volume}{158} (\bibinfo{year}{2019}), \bibinfo{pages}{110390}.
\newblock


\bibitem[Holstein et~al\mbox{.}(2019)]%
        {holstein2019improving}
\bibfield{author}{\bibinfo{person}{Kenneth Holstein}, \bibinfo{person}{Jennifer Wortman~Vaughan}, \bibinfo{person}{Hal Daum\'{e}}, \bibinfo{person}{Miro Dudik}, {and} \bibinfo{person}{Hanna Wallach}.} \bibinfo{year}{2019}\natexlab{}.
\newblock \showarticletitle{Improving Fairness in Machine Learning Systems: What Do Industry Practitioners Need?}. In \bibinfo{booktitle}{\emph{Proceedings of the 2019 CHI Conference on Human Factors in Computing Systems}} (Glasgow, Scotland Uk) \emph{(\bibinfo{series}{CHI '19})}. \bibinfo{publisher}{Association for Computing Machinery}, \bibinfo{address}{New York, NY, USA}, \bibinfo{pages}{1–16}.
\newblock
\showISBNx{9781450359702}
\urldef\tempurl%
\url{https://doi.org/10.1145/3290605.3300830}
\showDOI{\tempurl}


\bibitem[Ilyas and Khan(2017)]%
        {ilyas2017empirical}
\bibfield{author}{\bibinfo{person}{Muhammad Ilyas} {and} \bibinfo{person}{Siffat~Ullah Khan}.} \bibinfo{year}{2017}\natexlab{}.
\newblock \showarticletitle{An empirical investigation of the software integration success factors in GSD environment}. In \bibinfo{booktitle}{\emph{2017 IEEE 15th International Conference on Software Engineering Research, Management and Applications (SERA)}}. \bibinfo{publisher}{IEEE}, \bibinfo{address}{London, United Kingdom}, \bibinfo{pages}{255--262}.
\newblock


\bibitem[Khan et~al\mbox{.}(2021)]%
        {khan2021critical}
\bibfield{author}{\bibinfo{person}{Shams~Ullah Khan}, \bibinfo{person}{Abudul~Wahid Khan}, \bibinfo{person}{Faheem Khan}, \bibinfo{person}{Muhammad~Adnan Khan}, {and} \bibinfo{person}{Taeg~Keun Whangbo}.} \bibinfo{year}{2021}\natexlab{}.
\newblock \showarticletitle{Critical success factors of component-based software outsourcing development from vendors’ perspective: a systematic literature review}.
\newblock \bibinfo{journal}{\emph{IEEE Access}}  \bibinfo{volume}{10} (\bibinfo{year}{2021}), \bibinfo{pages}{1650--1658}.
\newblock


\bibitem[Kitchenham and Pfleeger(2008)]%
        {kitchenham2008personal}
\bibfield{author}{\bibinfo{person}{Barbara~A Kitchenham} {and} \bibinfo{person}{Shari~L Pfleeger}.} \bibinfo{year}{2008}\natexlab{}.
\newblock \showarticletitle{Personal opinion surveys}.
\newblock In \bibinfo{booktitle}{\emph{Guide to advanced empirical software engineering}}. \bibinfo{publisher}{Springer}, \bibinfo{address}{Berlin, Germany}, \bibinfo{pages}{63--92}.
\newblock


\bibitem[Konys et~al\mbox{.}(2013)]%
        {konys2013approach}
\bibfield{author}{\bibinfo{person}{Agnieszka Konys}, \bibinfo{person}{Jaros{\l}aw W{\k{a}}tr{\'o}bski}, {and} \bibinfo{person}{Przemys{\l}aw R{\'o}{\.z}ewski}.} \bibinfo{year}{2013}\natexlab{}.
\newblock \showarticletitle{Approach to practical ontology design for supporting COTS component selection processes}. In \bibinfo{booktitle}{\emph{Intelligent Information and Database Systems: 5th Asian Conference, ACIIDS 2013, Kuala Lumpur, Malaysia, March 18-20, 2013, Proceedings, Part II 5}}. \bibinfo{publisher}{Springer}, \bibinfo{address}{Kuala Lumpur, Malaysia}, \bibinfo{pages}{245--255}.
\newblock


\bibitem[Kumar and Bhatia(2015)]%
        {kumar2015neuro}
\bibfield{author}{\bibinfo{person}{Gaurav Kumar} {and} \bibinfo{person}{Pradeep~Kumar Bhatia}.} \bibinfo{year}{2015}\natexlab{}.
\newblock \showarticletitle{Neuro-Fuzzy model to estimate \& optimize quality and performance of component based software engineering}.
\newblock \bibinfo{journal}{\emph{ACM SIGSOFT Software Engineering Notes}} \bibinfo{volume}{40}, \bibinfo{number}{2} (\bibinfo{year}{2015}), \bibinfo{pages}{1--6}.
\newblock


\bibitem[Land et~al\mbox{.}(2008)]%
        {land2008cots}
\bibfield{author}{\bibinfo{person}{Rikard Land}, \bibinfo{person}{Laurens Blankers}, \bibinfo{person}{Michel Chaudron}, {and} \bibinfo{person}{Ivica Crnkovi{\'c}}.} \bibinfo{year}{2008}\natexlab{}.
\newblock \showarticletitle{COTS selection best practices in literature and in industry}. In \bibinfo{booktitle}{\emph{International Conference on Software Reuse}}. \bibinfo{publisher}{Springer}, \bibinfo{address}{Bejing, China}, \bibinfo{pages}{100--111}.
\newblock


\bibitem[Larios~Vargas et~al\mbox{.}(2020)]%
        {larios2020selecting}
\bibfield{author}{\bibinfo{person}{Enrique Larios~Vargas}, \bibinfo{person}{Maur\'{\i}cio Aniche}, \bibinfo{person}{Christoph Treude}, \bibinfo{person}{Magiel Bruntink}, {and} \bibinfo{person}{Georgios Gousios}.} \bibinfo{year}{2020}\natexlab{}.
\newblock \showarticletitle{Selecting third-party libraries: the practitioners’ perspective}. In \bibinfo{booktitle}{\emph{Proceedings of the 28th {ACM} Joint Meeting on European Software Engineering Conference and Symposium on the Foundations of Software Engineering}} (Virtual Event, USA) \emph{(\bibinfo{series}{ESEC/FSE 2020})}. \bibinfo{publisher}{Association for Computing Machinery}, \bibinfo{address}{New York, NY, USA}, \bibinfo{pages}{245–256}.
\newblock
\showISBNx{9781450370431}
\urldef\tempurl%
\url{https://doi.org/10.1145/3368089.3409711}
\showDOI{\tempurl}


\bibitem[Lenarduzzi et~al\mbox{.}(2020)]%
        {lenarduzzi2020open}
\bibfield{author}{\bibinfo{person}{Valentina Lenarduzzi}, \bibinfo{person}{Davide Taibi}, \bibinfo{person}{Davide Tosi}, \bibinfo{person}{Luigi Lavazza}, {and} \bibinfo{person}{Sandro Morasca}.} \bibinfo{year}{2020}\natexlab{}.
\newblock \showarticletitle{Open source software evaluation, selection, and adoption: a systematic literature review}. In \bibinfo{booktitle}{\emph{2020 46th Euromicro Conference on Software Engineering and Advanced Applications (SEAA)}}. \bibinfo{publisher}{IEEE}, \bibinfo{address}{Portorož, Slovenia}, \bibinfo{pages}{437--444}.
\newblock


\bibitem[Li et~al\mbox{.}(2005)]%
        {li2005empirical}
\bibfield{author}{\bibinfo{person}{Jingyue Li}, \bibinfo{person}{Reidar Conradi}, \bibinfo{person}{Odd Petter~N Slyngstad}, \bibinfo{person}{Christian Bunse}, \bibinfo{person}{Umair Khan}, \bibinfo{person}{Marco Torchiano}, {and} \bibinfo{person}{Maurizio Morisio}.} \bibinfo{year}{2005}\natexlab{}.
\newblock \showarticletitle{An empirical study on off-the-shelf component usage in industrial projects}. In \bibinfo{booktitle}{\emph{Product Focused Software Process Improvement: 6th International Conference, PROFES 2005, Oulu, Finland, June 13-15, 2005. Proceedings 6}}. \bibinfo{publisher}{Springer}, \bibinfo{address}{Oulu, Finland}, \bibinfo{pages}{54--68}.
\newblock


\bibitem[Li and Sun(2023)]%
        {li2023challenges}
\bibfield{author}{\bibinfo{person}{Yi Li} {and} \bibinfo{person}{Meng Sun}.} \bibinfo{year}{2023}\natexlab{}.
\newblock \showarticletitle{Challenges Engaging Formal CBSE in Industrial Applications}. In \bibinfo{booktitle}{\emph{International Conference on Formal Aspects of Component Software}}. \bibinfo{publisher}{Springer Nature Switzerland}, \bibinfo{address}{Cham}, \bibinfo{pages}{153--167}.
\newblock


\bibitem[Mahapatro and Padhy(2024)]%
        {mahapatro2024reviewing}
\bibfield{author}{\bibinfo{person}{Pradeep~Kumar Mahapatro} {and} \bibinfo{person}{Neelamadhab Padhy}.} \bibinfo{year}{2024}\natexlab{}.
\newblock \showarticletitle{Reviewing the Landscape: Component-Based Software Engineering Practices and Challenges}. In \bibinfo{booktitle}{\emph{2024 International Conference on Emerging Systems and Intelligent Computing (ESIC)}}. \bibinfo{publisher}{IEEE}, \bibinfo{address}{Bhubaneswar, India}, \bibinfo{pages}{360--365}.
\newblock


\bibitem[Nabot(2024)]%
        {nabot2024software}
\bibfield{author}{\bibinfo{person}{Ahmad Nabot}.} \bibinfo{year}{2024}\natexlab{}.
\newblock \showarticletitle{Software component selection methods and techniques: a systematic review}.
\newblock \bibinfo{journal}{\emph{Indonesian Journal of Electrical Engineering and Computer Science}} \bibinfo{volume}{33}, \bibinfo{number}{3} (\bibinfo{year}{2024}), \bibinfo{pages}{1802--1811}.
\newblock


\bibitem[Nadi and Sakr(2023)]%
        {nadi2023selecting}
\bibfield{author}{\bibinfo{person}{Sarah Nadi} {and} \bibinfo{person}{Nourhan Sakr}.} \bibinfo{year}{2023}\natexlab{}.
\newblock \showarticletitle{Selecting third-party libraries: the data scientist’s perspective}.
\newblock \bibinfo{journal}{\emph{Empirical Software Engineering}} \bibinfo{volume}{28}, \bibinfo{number}{1} (\bibinfo{year}{2023}), \bibinfo{pages}{15}.
\newblock
\urldef\tempurl%
\url{https://doi.org/10.1007/s10664-022-10241-3}
\showDOI{\tempurl}


\bibitem[Petersen et~al\mbox{.}(2017)]%
        {petersen2017choosing}
\bibfield{author}{\bibinfo{person}{Kai Petersen}, \bibinfo{person}{Deepika Badampudi}, \bibinfo{person}{Syed Muhammad~Ali Shah}, \bibinfo{person}{Krzysztof Wnuk}, \bibinfo{person}{Tony Gorschek}, \bibinfo{person}{Efi Papatheocharous}, \bibinfo{person}{Jakob Axelsson}, \bibinfo{person}{Severine Sentilles}, \bibinfo{person}{Ivica Crnkovic}, {and} \bibinfo{person}{Antonio Cicchetti}.} \bibinfo{year}{2017}\natexlab{}.
\newblock \showarticletitle{Choosing component origins for software intensive systems: In-house, COTS, OSS or outsourcing?—A case survey}.
\newblock \bibinfo{journal}{\emph{IEEE Transactions on Software Engineering}} \bibinfo{volume}{44}, \bibinfo{number}{3} (\bibinfo{year}{2017}), \bibinfo{pages}{237--261}.
\newblock


\bibitem[Santos et~al\mbox{.}(2019)]%
        {santos2019mind}
\bibfield{author}{\bibinfo{person}{Ronnie~ES Santos}, \bibinfo{person}{Ay{\c{s}}e Bener}, \bibinfo{person}{Maria~Teresa Baldassarre}, \bibinfo{person}{Cleyton~VC Magalh{\~a}es}, \bibinfo{person}{Jorge~S Correia-Neto}, {and} \bibinfo{person}{Fabio~QB da Silva}.} \bibinfo{year}{2019}\natexlab{}.
\newblock \showarticletitle{Mind the gap: are practitioners and researchers in software testing speaking the same language?}. In \bibinfo{booktitle}{\emph{2019 IEEE/ACM Joint 7th International Workshop on Conducting Empirical Studies in Industry (CESI) and 6th International Workshop on Software Engineering Research and Industrial Practice (SER\&IP)}}. \bibinfo{publisher}{IEEE}, \bibinfo{address}{Montreal, QC, Canada}, \bibinfo{pages}{10--17}.
\newblock


\bibitem[Shull et~al\mbox{.}(2008)]%
        {shull2008guide}
\bibfield{author}{\bibinfo{person}{Forrest Shull}, \bibinfo{person}{Janice Singer}, {and} \bibinfo{person}{Dag~IK Sj{\o}berg}.} \bibinfo{year}{2008}\natexlab{}.
\newblock \bibinfo{booktitle}{\emph{Guide to advanced empirical software engineering}}. Vol.~\bibinfo{volume}{93}.
\newblock \bibinfo{publisher}{Springer}, \bibinfo{address}{Berlin, Germany}.
\newblock


\bibitem[Siena et~al\mbox{.}(2014)]%
        {siena2014modelling}
\bibfield{author}{\bibinfo{person}{Alberto Siena}, \bibinfo{person}{Mirko Morandini}, {and} \bibinfo{person}{Angelo Susi}.} \bibinfo{year}{2014}\natexlab{}.
\newblock \showarticletitle{Modelling risks in open source software component selection}. In \bibinfo{booktitle}{\emph{Conceptual Modeling: 33rd International Conference, ER 2014, Atlanta, GA, USA, October 27-29, 2014. Proceedings 33}}. \bibinfo{publisher}{Springer}, \bibinfo{address}{Atlanta, GA, USA}, \bibinfo{pages}{335--348}.
\newblock


\bibitem[Storey et~al\mbox{.}(2020)]%
        {storey2020software}
\bibfield{author}{\bibinfo{person}{Margaret-Anne Storey}, \bibinfo{person}{Neil~A Ernst}, \bibinfo{person}{Courtney Williams}, {and} \bibinfo{person}{Eirini Kalliamvakou}.} \bibinfo{year}{2020}\natexlab{}.
\newblock \showarticletitle{The who, what, how of software engineering research: a socio-technical framework}.
\newblock \bibinfo{journal}{\emph{Empirical Software Engineering}}  \bibinfo{volume}{25} (\bibinfo{year}{2020}), \bibinfo{pages}{4097--4129}.
\newblock


\bibitem[Vale et~al\mbox{.}(2016)]%
        {vale2016twenty}
\bibfield{author}{\bibinfo{person}{Tassio Vale}, \bibinfo{person}{Ivica Crnkovic}, \bibinfo{person}{Eduardo~Santana De~Almeida}, \bibinfo{person}{Paulo Anselmo da Mota~Silveira Neto}, \bibinfo{person}{Yguarat{\~a}~Cerqueira Cavalcanti}, {and} \bibinfo{person}{Silvio~Romero de Lemos~Meira}.} \bibinfo{year}{2016}\natexlab{}.
\newblock \showarticletitle{Twenty-eight years of component-based software engineering}.
\newblock \bibinfo{journal}{\emph{Journal of Systems and Software}}  \bibinfo{volume}{111} (\bibinfo{year}{2016}), \bibinfo{pages}{128--148}.
\newblock


\bibitem[Wohlin(2013)]%
        {wohlin2013empirical}
\bibfield{author}{\bibinfo{person}{Claes Wohlin}.} \bibinfo{year}{2013}\natexlab{}.
\newblock \showarticletitle{Empirical software engineering research with industry: Top 10 challenges}. In \bibinfo{booktitle}{\emph{2013 1st international workshop on conducting empirical studies in industry (CESI)}}. \bibinfo{publisher}{IEEE}, \bibinfo{address}{San Francisco, CA, USA}, \bibinfo{pages}{43--46}.
\newblock


\bibitem[Wohlin et~al\mbox{.}(2012)]%
        {wohlin2012experimentation}
\bibfield{author}{\bibinfo{person}{Claes Wohlin}, \bibinfo{person}{Per Runeson}, \bibinfo{person}{Martin H{\"o}st}, \bibinfo{person}{Magnus~C Ohlsson}, \bibinfo{person}{Bj{\"o}rn Regnell}, \bibinfo{person}{Anders Wessl{\'e}n}, {et~al\mbox{.}}} \bibinfo{year}{2012}\natexlab{}.
\newblock \bibinfo{booktitle}{\emph{Experimentation in software engineering}}. Vol.~\bibinfo{volume}{236}.
\newblock \bibinfo{publisher}{Springer}, \bibinfo{address}{Heidelberg, Germany}.
\newblock


\end{thebibliography}

\end{document}